\documentclass[12pt]{article}

\usepackage{amsmath}
\usepackage{amsfonts}
\usepackage{amssymb}
\usepackage{latexsym}
\usepackage{graphicx}
\usepackage{textcomp}
\usepackage[authoryear]{natbib}
\usepackage[utf8]{inputenc}
\usepackage[activeacute,english]{babel}
\newcommand\etal{\mbox{\textit{et al.}}}

\begin{document}

\title{ The dengue fever in four municipalities of the Colima state, Mexico, from 2008 to 2015 and its relation with some climatological normals}

\author{ Erick J. L\'opez-S\'anchez $^{(1)}$\thanks{lsej@unam.mx}, Norma Y. S\'anchez-Torres$^{(2)}$\thanks{norma.sanchez\_torres@etu.upmc.fr}, \\
 Carlos Trenado$^{(3)}$\thanks{carlos.trenado@med.uni-duesseldorf.de}    and Juan M. Romero $^{(4)}$ \thanks{jromero@correo.cua.uam.mx}\\[0.5cm]
\it $^{(1)}$Posgrado en Ciencias Naturales e Ingenieria, \\
\it Universidad Aut\'onoma Metropolitana, Cuajimalpa. \\
\it Vasco de Quiroga 4871, Santa Fe Cuajimalpa, D. F. 05300 M\'exico.\\
\it$^{(2)}$ Universit\'e Pierre et Marie CURIE, 4 Place Jussieu 75005 Paris, France. \\
\it $^{(3)}$Medical Faculty, Heinrich-Heine-University D\"{u}sseldorf, \\
 \it Moorenstra{\ss}e 5 40225 D\"{u}sseldorf, Germany \\
 \it $^{(4)}$Departamento de Matem\'aticas Aplicadas y Sistemas,\\
\it Universidad Aut\'onoma Metropolitana-Cuajimalpa,\\
\it M\'exico, D.F  05300, M\'exico\\
}

\pagestyle{plain}
\date{}

\maketitle

\begin {abstract}
Although dengue fever is an ancient disease that  should have been eradicated since time ago, in Mexico as well as in other regions worldwide the cases of classic and hemorrhagic dengue continue to be reported presently. Since 2002 (and before), a worrying outbreak of dengue was observed in several states of Mexico, notably the case of Colima  for which a {\it ``descacharrizaci\'on''} campaign was implemented in 2009 to encourage the population to get rid of old objects accumulated in their backyards and outdoors so as to prevent proliferation of mosquitoes {\it Aedes Aegypti}. To understand the effect of such campaign, we studied the incidence of dengue in four municipalities of the mentioned state, namely Manzanillo, Armer\'ia, Tecom\'an and Ixtlahuac\'an. In particular, we observed that the incidence of dengue in these four municipalities had a strong relationship with climatological normals, specifically precipitation and minimum temperature. Furthermore, the results showed that the mentioned campaign helped prevent mosquito eggs incubation and their development, thus leading to reduction of the incidence of dengue in the human population.

\end{abstract}

\section{Introduction}  \label{intro}

Dengue is a vector-borne disease that is transferred by the female mosquito {\it Aedes Aegypti}, who by simply taking blood from a sick person is capable of passing the virus throughout their live including their descendants \citep{Ha08}. These mosquitoes live worldwide between the latitudes 35\textdegree N and 35\textdegree S, reproduce at temperatures between 25\textcelsius-30\textcelsius \ and are still able to survive at temperatures over 10 \textcelsius \  \citep{De09, Ei14}. In addition, mosquitoes are attracted by CO$_{2}$ \citep{Mu92, Cu12} and their average lifespan of two weeks  may increase in warm or moist environments \citep{Ma15}.  \\

Dengue fever is commonly known as a tropical disease, however very recent reports emphasize the effect of climate change in dengue spreading while \citet{Ca13} also stressing its incidence in new geographic locations noteworthy the case of Portugal (Madeira), France, Croatia, United States of America (Florida) and China (Yuan) \citep{Fa14,Fr15,Ma13,La10,Se14,Eu14,Ca13,Ra12,Ky08}.  In fact, dengue incidence has dramatically increased over the last 50 years according to reports in 128 countries \citep{Br12}. Thus, understanding and finding ways to eradicate dengue fever remains a critical task for local and global public health organizations. \\

Today the only method to cope with dengue fever is by controlling the spread of mosquitoes {\it Aedes Aegypti} as there is currently no effective vaccine. For this purpose, the World Health Organization (WHO) recommends actions such as environmental management of mosquito populations by proper disposal of solid waste and removal of artificial water containers, covering domestic water storage containers, window screens, long-sleeved clothes,  coils and vaporizers \citep{Wh14}; chemical control of dengue mosquitoes by application of appropriate insecticides to outdoor water storage containers, insecticide-treated materials, repellents \citep{Wo09}. Novel control approaches envision the use of biological control agents for instance predatory crustaceans aimed at prompt elimination of colonization foci and the release of genetically engineered mosquito populations with reduced lifespan \citep{Mc09}.  \\

In Mexico the {\it Aedes Aegypti} mosquito has been found in states like Nuevo Le\'on, Coahuila, Tamaulipas, Veracruz, Colima and Chiapas \citep{Sa12,Lo12, Na15}. In these states numerous cases of classic  and hemorrhagic dengue continue to be reported. Notably, the case of Colima which was the state with the highest number of dengue cases reported in 2002 \citep{Da14, Ch02}. \\

The state of Colima in Mexico possesses physical and geographical characteristics leading to an average temperature between 18\textcelsius \ and 28\textcelsius, which facilitates the development of {\it Aedes Aegypti} mosquitoes. Under such climatological conditions, the state of Colima had a historical maximum of 9630 cases in 2002. This phenomenon not only led to population's suffering but also economic burden for the government and society. To cope with this, the state government emphasized research investment and at first stage several prevention actions involving community participation as well as larval and chemical control that was supported by epidemiological and entomological surveillance. Further successful efforts included the ``descharrizaci\'on'' campaign that attained very positive results as shown in the present report. \\

In this paper, we aim at establishing a simple qualitative relationship between the periodicity of temperature and precipitation in relation to dengue incidence in four municipalities of the state of Colima (Manzanillo, Armer\'ia, Tecom\'an and Ixtlahuac\'an). To this end, we targeted the analysis of data concerning precipitation, minimum temperature and dengue cases reported between the years 2008 and 2015. It is concluded that the reported relationship and methodology may be useful to define predictive patterns of dengue occurrence that support public health decision-making in the case of coastal and isolated municipalities.\\


\section{Background} \label{se:Antecedentes}

\begin{figure}[t!] 
\begin{center}
\centerline{\includegraphics[width=1\textwidth]{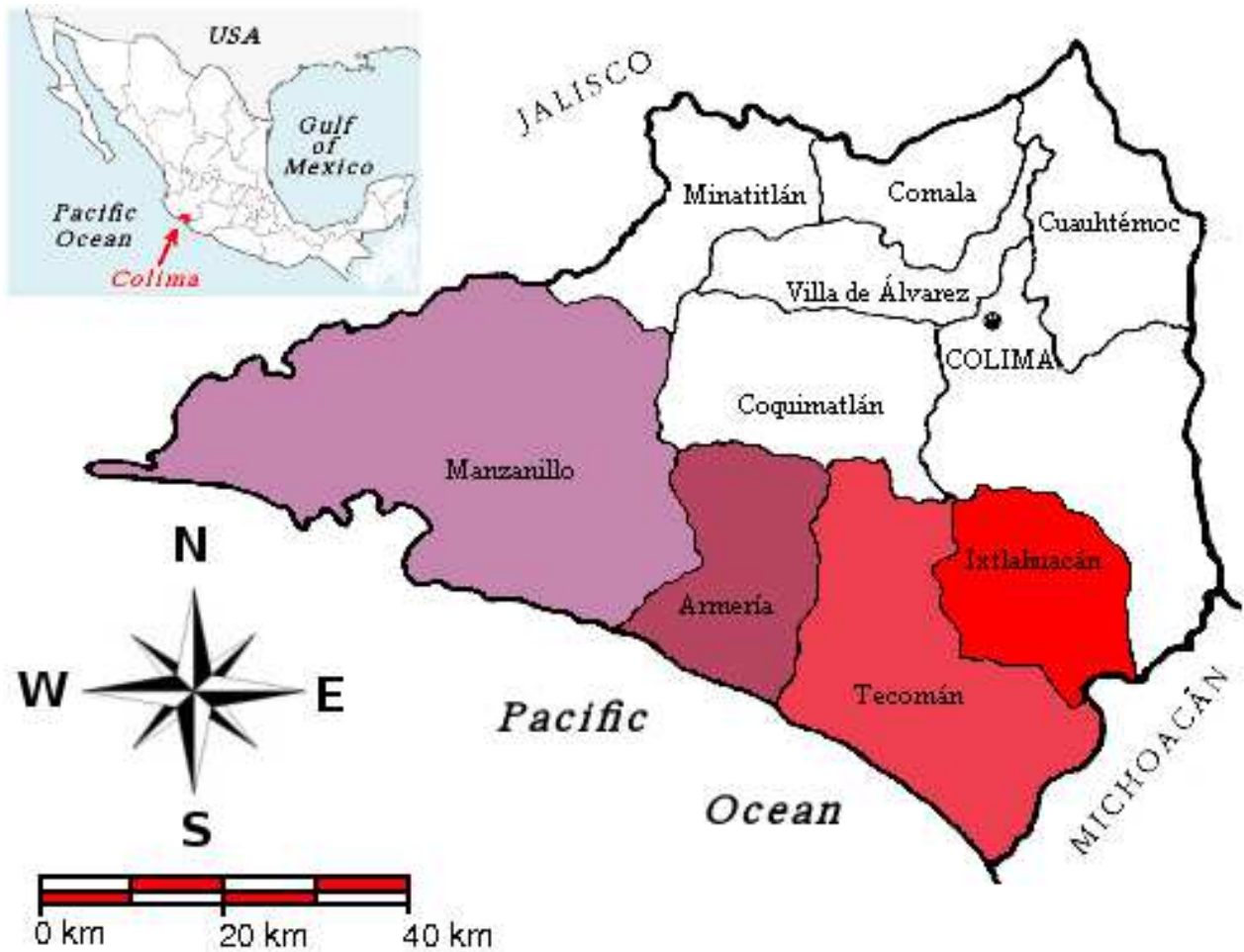}}
\caption{{\small Municipalities most vulnerable to the impact of dengue in the period 2008-2014. Modified from \cite{Mcrd,Me15}. }} \label{mapa}
\end{center}
\end{figure}

Colima is a state possessing physical and geographical characteristics that are optimal for dengue mosquito development, namely a temperature between 18\textcelsius \ and 28\textcelsius. It is located in a geographical position between the coordinates of 19\textdegree05'48'' north latitude and 103\textdegree57'39'' west longitude. A map of Colima depicting the location of its municipalities is shown in Fig. \ref{mapa}. \\
 
Since many years the population of Colima has suffered from the presence of the mosquito {\it Aedes aegypti} and consequently dengue fever. Noteworthy, a historical maximum of 9630 cases was reported in 2002, which translated into population's suffering and considerable economic loss for the state due to high costs of hospitalization. For this reason the state government decided to implement actions at the heart of public policy, firstly by investing in research and prevention through community participation, larval control and chemical control, supported by epidemiological and entomological surveillance; secondly by launching in 2009 the {\it descacharrizaci\'on} campaign whose slogan was: ``Everyone was together at the same time, the same task: removing pots'' \citep{An11}.  \\

The Prevention and Control Diseases National Center [CENAPRECE, Mexican National Organization for Control of Disasters, \citep{Na15}] promoted strategies that included early action to reduce the risks of transmission, hold back dengue outbreaks and prevent their spread as a quick response for the occurrence of infections. CENAPRECE further requested their strategies to support social programs and community participation involving all three levels of government, namely federal, state and municipal levels. \\

Preventive measures in the Colima state to cope with dengue began in 2010, just after Colima was reported as the state with the highest incidence of dengue at the national level in 2009. The program was conducted under the auspices of the University of Colima \citep{Uc09}, which was granted funds by CONACYT (Mexican Science Foundation). On the other hand,  \citet{Sx14} allocated the Government of Colima three million Mexican pesos to fight the disease, while the state government earmarked 14 million pesos for insecticides and programs to prevent and combat dengue (in Diario de Colima, October 2009).  \\

The aim of the program was to reduce the infection risks, keep an epidemiological control and minimize the number of deaths caused by dengue in the state. To achieve this, the implemented strategies included spray insecticide on schools and public spaces, allocation of one ovitrap for each group of 30 houses, etc. \citep{An11}. As reported, outbreaks occurred when the temperature increased and due to precipitations \citep{Dc13}.  \\

\section{ Materials and methods } \label{se:metodologia}

\citet{He12} studied a relationship between dengue incidence and temperature in order to predict disease hot spots. The methodology that we utilize in the present report is inspired by this previous study. In particular, we realized that dengue incidence occurs when local maxima of the minimal temperature curve coincide with precipitation. \\

The maximum and minimum temperature curves and pluvial data were obtained from the Water National Commission \citep{Cm14}. The periodicity of such data is daily. The classical dengue incidence data were taken from the bulletin of Epidemiology General Direction \citep{Bo14}. The periodicity of incidence data is weekly.  \\

\subsection{ Data treatment}

\begin{figure}[hbt] 
\begin{center}
\centerline{\includegraphics[width=1\textwidth]{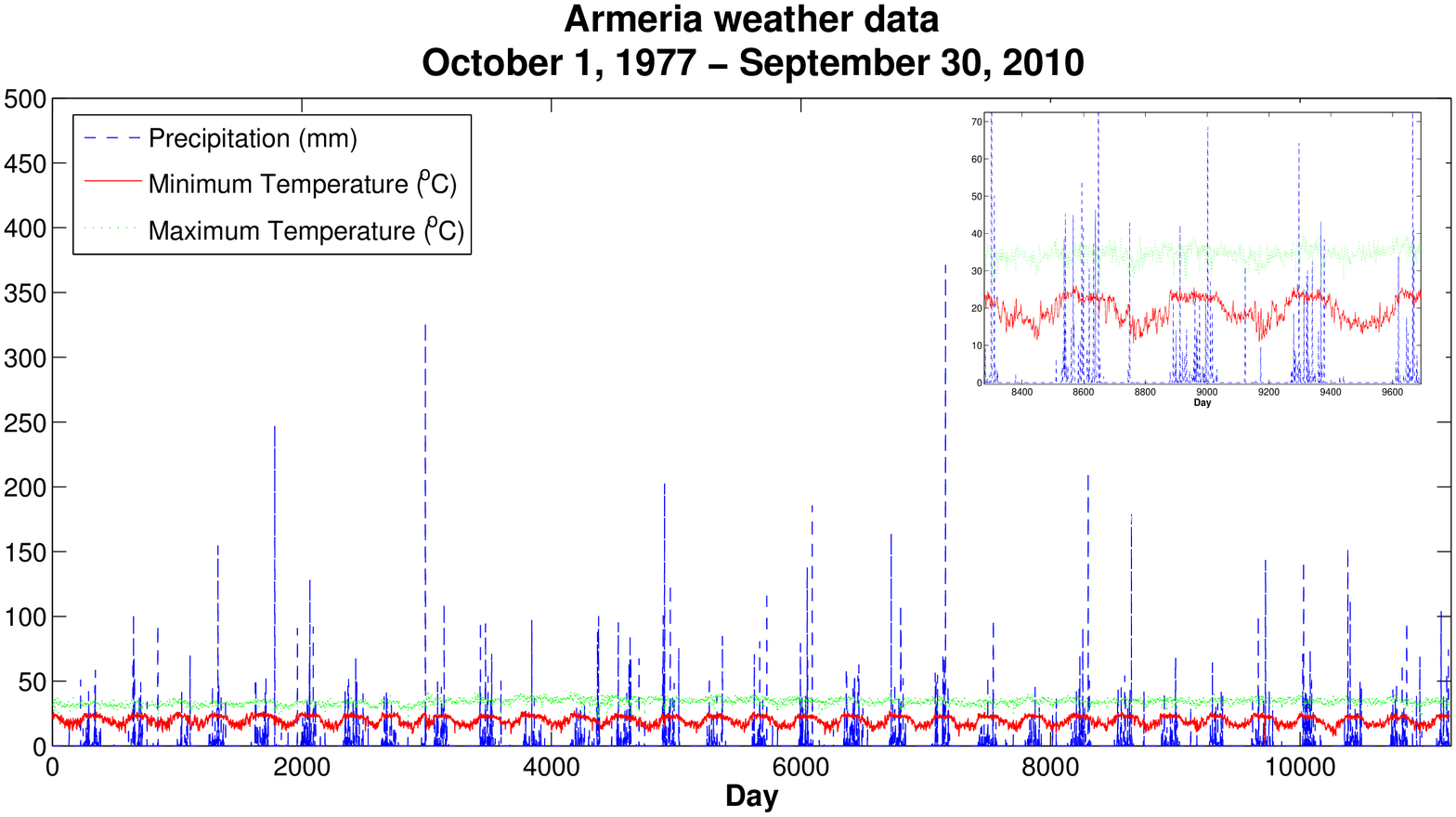}}
\caption{{\small Three meteorological varables for Armer\'ia, Colima.}} \label{arme_meteo}
\end{center}
\end{figure}

The CNA provides weather data with daily periodicity from 1949 to 2014. 
These data show a periodic behavior (see Fig. \ref{arme_meteo}). In order to verify such statement, the fast Fourier transform (FFT) was calculated on the obtained time series. Unfortunately, data provided by the CNA are incomplete, so it was necessary to perform cubic interpolations to fill the gaps of missing data.  \\

In Fig. \ref{arme_meteo} three meteorological variables are shown, namely minimum temperature, maximum temperature and precipitation for the Armer\'ia municipality. First day is October 1st, 1977. As noticeable, the minimum temperature variable and the precipitation have a periodic behavior. The same does not occur for the maximum temperature variable. We can see that precipitation occurred during maxima of the minimum temperature's curve. In the same figure a zoom of the four cycles is displayed.  \\

As epidemiological data available have weekly periodicity, it was necessary to obtain metereological data also with the same periodicity by calculating the mean of daily data. Epidemiological data were taken from January 2008 until June 2013. This time interval includes the {\it descacharrizaci\'on} campaign conducted between 2009 and 2010. We exploited the fact about the periodicity of the meteorological data to extrapolate data for the weeks that were not available, i. e. from June 2013 until June 2015, that is to say data from week 364 until week 440.  \\

\begin{table}[ht]
  \begin{center}

  \begin{tabular}{cc|cc}
\hline 
{\bf Year} & {\it Week} & {\bf Year} & {\it Week}  \\ 
\hline
2007        & 1 to 52 & 2012	       & 261 to 312   \\
2008	       & 53 to 104 & 2013   &  313 to 364 \\
2009 	       & 105 to 156 & 2014   & 365 to 416   \\
2010	       & 157 to 208 & 2015   &  417 to 468 \\
2011	       & 209 to 260  & 2016  & 469 to 520 \\  
\hline
   \end{tabular}
  \caption{Distribution of the weeks by year.} \label{cuadro}
  \label{tab:kd}
  \end{center}
\end{table}

It is important to note that the meteorological variables had a similar behavior in all municipalities that we addressed, so we only show data for the Armer\'ia municipality to avoid redundancy. The first week was defined from January 1, 2007. See table \ref{cuadro} for details. \\

The CNA temperature and precipitation data are periodic [we test them by aplying the Fast Fourier Transform on the corresponding time series \citep{Si14}]. We use the variable minimum temperature because it is more stable than the variable maximum temperature.  \\

\section{Results} \label{se:results}

The red flag terminology that the Directorate General of Epidemiology (DGE) of the Ministry of health in Mexico assigns to a municipality, does not necessarily indicate occurrence of a big number of dengue incidences, but rather a sudden occurrence that demands immediate attention and monitoring. In particular, we focused on those four municipalities because they were listed as red flag in the period 2008-2013 by the DGE. \\

Weekly precipitation (mm), minimum temperature (\textcelsius) and number of infected persons (dengue incidence) are drawn for the four municipalities: Armer\'ia, Ixtlahuac\'an, Manzanillo and Tecom\'an.

\subsection{Armer\'ia} \label{se:armeria}

\begin{figure}[ht] 
\begin{center}
\centerline{\includegraphics[width=1\textwidth]{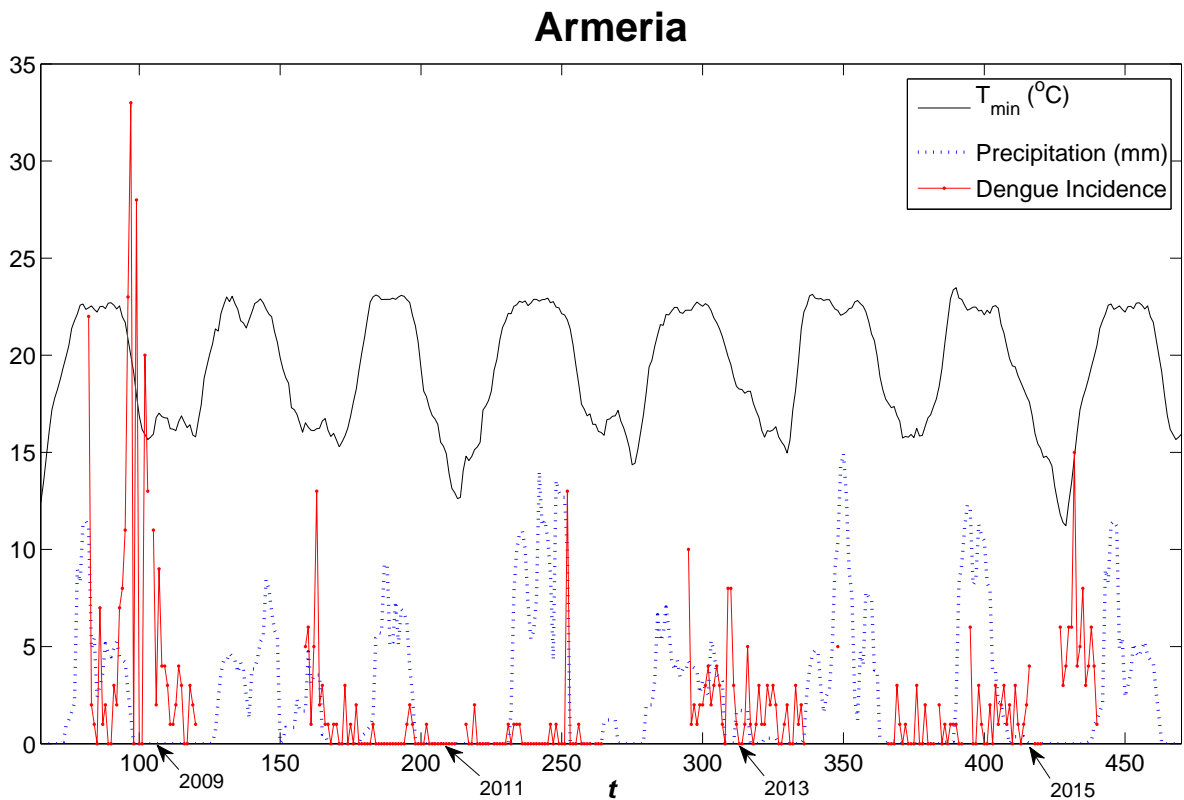}}
\caption{{\small Minimum temperature, precipitation and dengue incidence at Armer\'ia in de 2008-2015.}} \label{arme}
\end{center}
\end{figure}

Armer\'ia was one of the municipalities cataloged as red flag in 2009-2010, that is during the descacharrizacion campaign years. In Fig. \ref{arme}, it is noticeable that the last weeks of 2008 (around week 100) dengue incidence suddenly increased from 0 to 34 cases. This behavior caused Armer\'ia to be cataloged as red flag state and therefore it was continuously monitored to avoid a future epidemic.
In the year 2009 (week 105-156), the cases of dengue decreased until no more cases were reported for a period of about 30 weeks (week 125 to 155). \\

Nevertheless, new outbreaks occurred in the years 2010 and 2011, although the number of occurrences was lower than in 2008 (less or equal than 5 cases). Interestingly, the campaign of {\it descacharrizaci\'on} was taking place in other places at this time but not in Armer\'ia. Shortly after, a program consisting of prevention measures and fumigation was developed for this municipality. \\

Unfortunately, the year 2012 when the program was in its final stage, and in 2013 when the government decided not to allocate resources to control the disease, outbreaks of dengue emerged again, as seen in the period week 295-340  of Fig. \ref{arme}. Finally, the campaign of {\it descacharrizaci\'on} started in Armer\'ia in 2013 \citep{Ar13}. This year new dengue incidences occurred as observed in Fig. \ref{arme} and as can be observed, dengue incidence commonly followed a pattern of occurrence dictated by rainy seasons followed by temperature changes. However, in the last year (2014) outbreaks occurred without following such pattern, e.g. dengue cases were reported during a low value of the minimum temperature variable, before the next rainy season\footnote{Rains represented after week 450 in Fig. \ref{arme} have not occurred, they are only a prediction generated from the periodicity of the precipitation.}.\\

\subsection{Ixtlahuac\'an}  \label{se:ixtlahuacan}

Ixtlahuac\'an is one of the municipalities with the lowest population in the state of Colima, 5,300 people in 2010 [0.814689\% of the state population \citep{Sx14}]. However in 2009 it was listed as red flag \citep{Bo14}.  \\

\begin{figure}[ht] 
\begin{center}
\centerline{\includegraphics[width=1\textwidth]{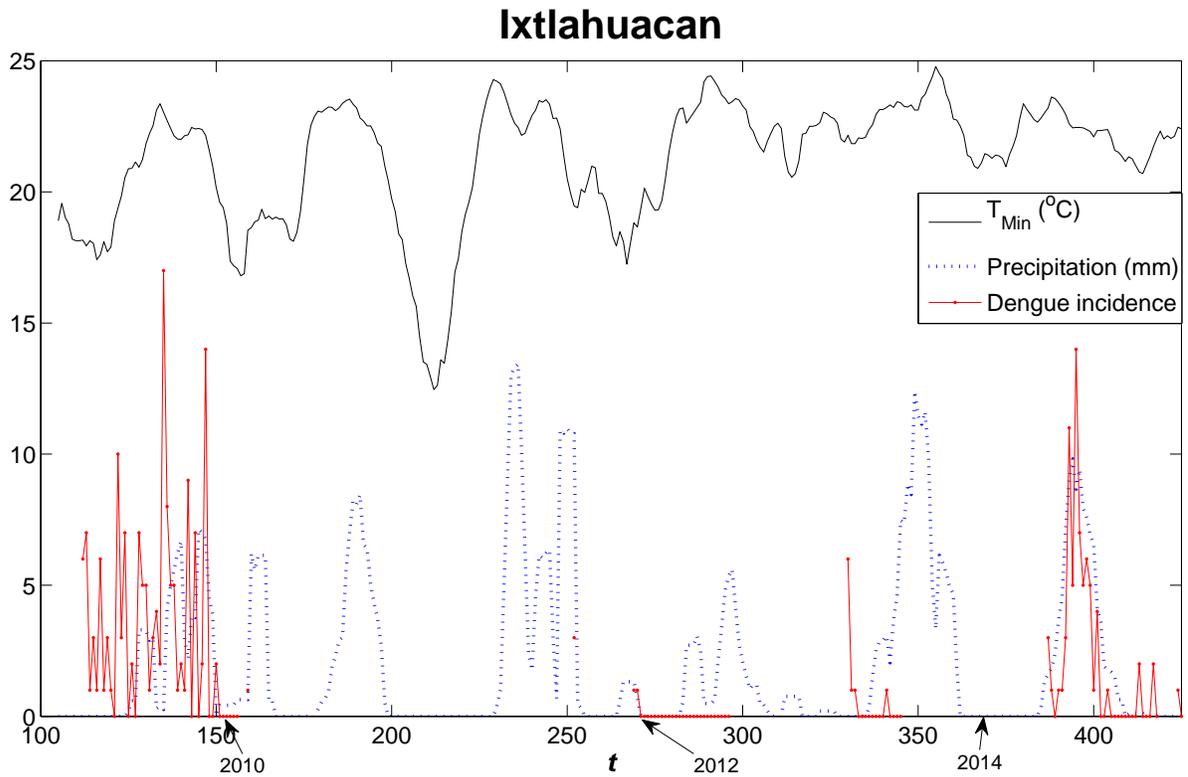}}
\caption{{\small Minimum temperature, precipitation and dengue incidence at Ixtahuacan in de 2008-2015.}} \label{ixtla}
\end{center}
\end{figure}

In Fig. \ref{ixtla}, the number of dengue cases reported and the weekly meteorological normals (minimum temperature and precipitation)  between the years 2009 and 2015 are displayed. Unlike the municipality of Armer\'ia, dengue incidence appears to follows a pattern characterized by rainy seasons and high values of the minimum temperature variable. The {\it descacharrizaci\'on} campaign started in 2009 in this municipality \citep{Cn14}, and had a positive effect as it can be seen in Fig. \ref{ixtla}, namely between 2010 and half of 2014 there was not an important number of dengue cases reported by the epidemiological bulletin. \\

Nevertheless, an important number of dengue cases were reported in the middle of 2014. By October 2014, the descacharrizacion campaign was implemented \citep{Ec14}. As observed in Fig. \ref{ixtla}, dengue incidences following the pattern of high values of minimum temperature and precipitation were observed again.

\subsection{Manzanillo}  \label{se:manzanillo}

\begin{figure}[ht] 
\begin{center}
\centerline{\includegraphics[width=1\textwidth]{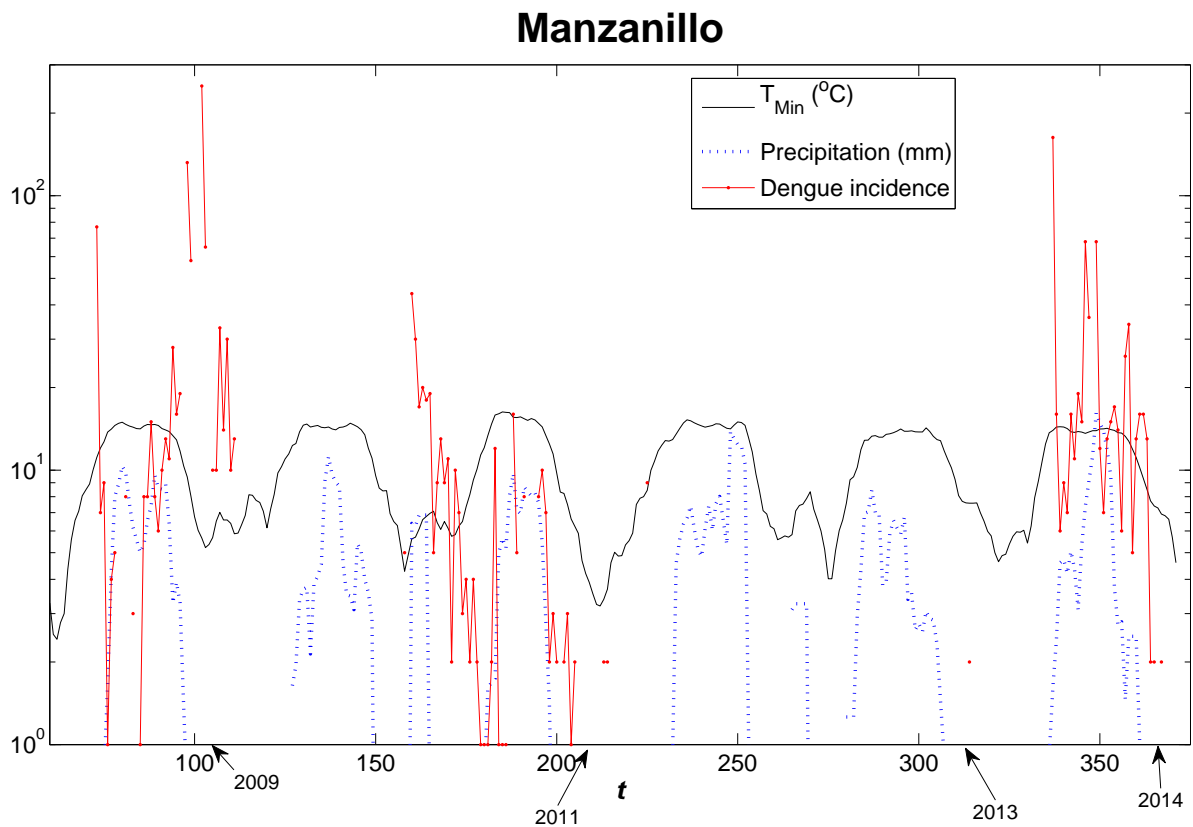}}
\caption{{\small Minimum temperature, precipitation and dengue incidence at Manzanillo in de 2008-2015.}} \label{manza}
\end{center}
\end{figure}

Manzanillo is one of the most important ports of Mexico. It is also one of the largest cities in the Colima state and all kind of visitors and goods arrive there by boat, plane, train, bus, etc. Importantly, the biggest number of dengue cases took place in Manzanillo during 2008 (see Fig. \ref{manza}). It is observed that most of the cases occurred in the years 2008 and 2009. \\

Dengue incidences ceased for a relatively long time (between the weeks 115 and 156) in 2009. However, dengue cases reappeared in 2010. As a result, the government implemented a new {\it descacharrizaci\'on} campaign. These campaigns helped to avoid new dengue outbreaks between 2011 and 2013. In those years there were only few dengue incidences reported, that is in the weeks 213 and 214 (two cases each), in the week 225 (eight cases) and in the week 314 (two cases). These few cases were not significant to consider Manzanillo as red flag state. \\

Dengue outbreaks started again when the descacharrizacion campaign finished in 2013. More than 151 confirmed dengue cases in Manzanillo (0.001\% of the total municipality population) were reported. As in Ixtlahuac\'an, the dengue incidences followed a pattern of occurrence guided by rains and with high values of the minimum temperature variable. As can be seen in Fig. \ref{manza}, only at the beginning of 2009 dengue cases occurred during low values of the minimum temperature variable and without precipitation.  \\

\subsection{Tecom\'an}   \label{se:tecoman} 

\begin{figure}[ht] 
\begin{center}
\centerline{\includegraphics[width=1\textwidth]{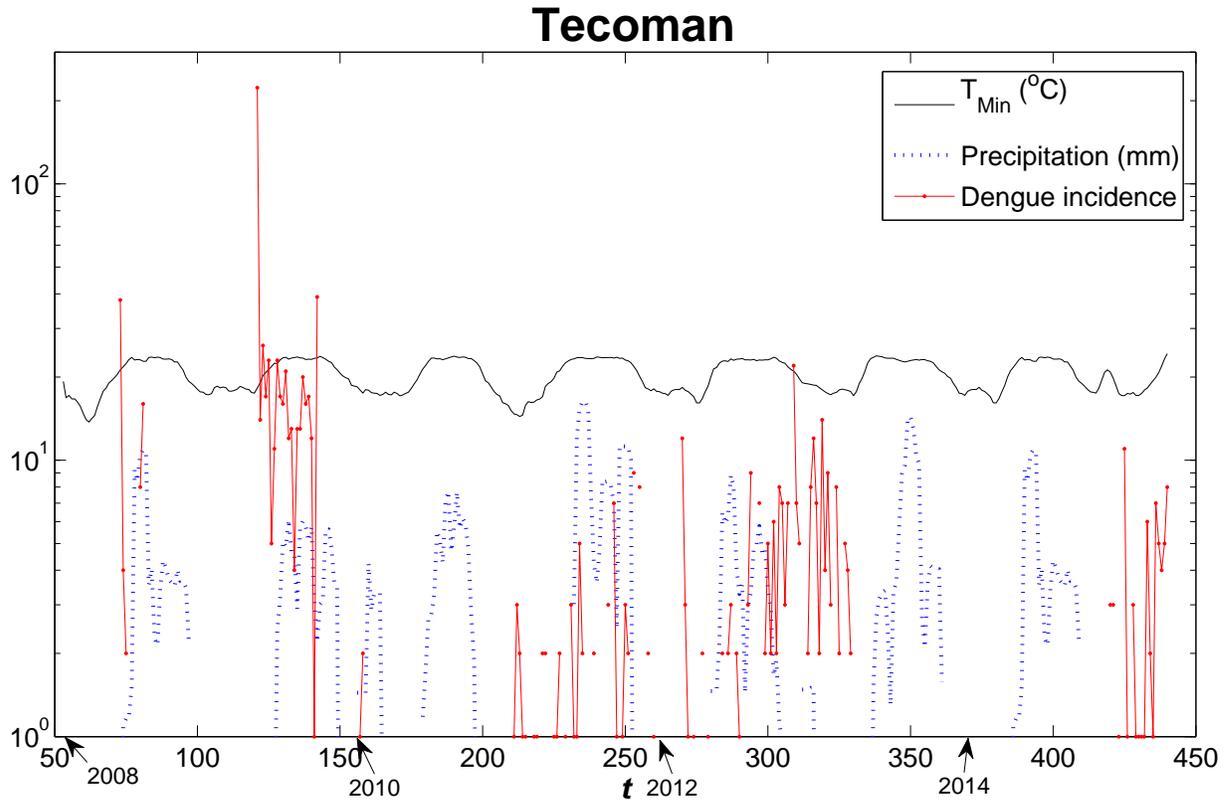}}
\caption{{\small Minimum temperature, precipitation and dengue incidence at Tecom\'an in de 2008--2015.}} \label{teco}
\end{center}
\end{figure}

Another coastal municipality of Colima is Tecom\'an. This municipality was cataloged as red flag between 2008 and 2009 because of the high incidence of dengue cases reported. In particular, Fig. \ref{teco} shows the incidence of dengue in Tecom\'an since 2008. It is noticeable that 2008 was not the year with more infected people with dengue, but rather 2009 with 220 cases approximately (0.02\% of the total municipality population). Throughout that year dengue was controlled and its occurrence decreased. Nevertheless during the last period of that year more than 130 confirmed cases had occurred.  \\

There were not reported cases of dengue in 2010, but in 2011 and 2012 the DGE assigned Tecom\'an red flag status, because of the occurrence of sudden outbreaks. Again, such outbreaks followed an occurrence pattern characterized by high values of the minimum temperature and precipitations. \\

The descacharrizacion campaign and other preventive measures helped mitigate the occurrence of dengue cases and in 2014 no dengue cases were reported. Nonetheless in 2015 there were new reports. Contrary to the occurrence pattern presented so far, in 2015 cases were reported during low values of the minimum temperature and without precipitation.

\section{Dengue fever in the 4 municipalities}  \label{se:compara}

\begin{figure}[hbt] 
\begin{center}
\centerline{\includegraphics[width=1\textwidth]{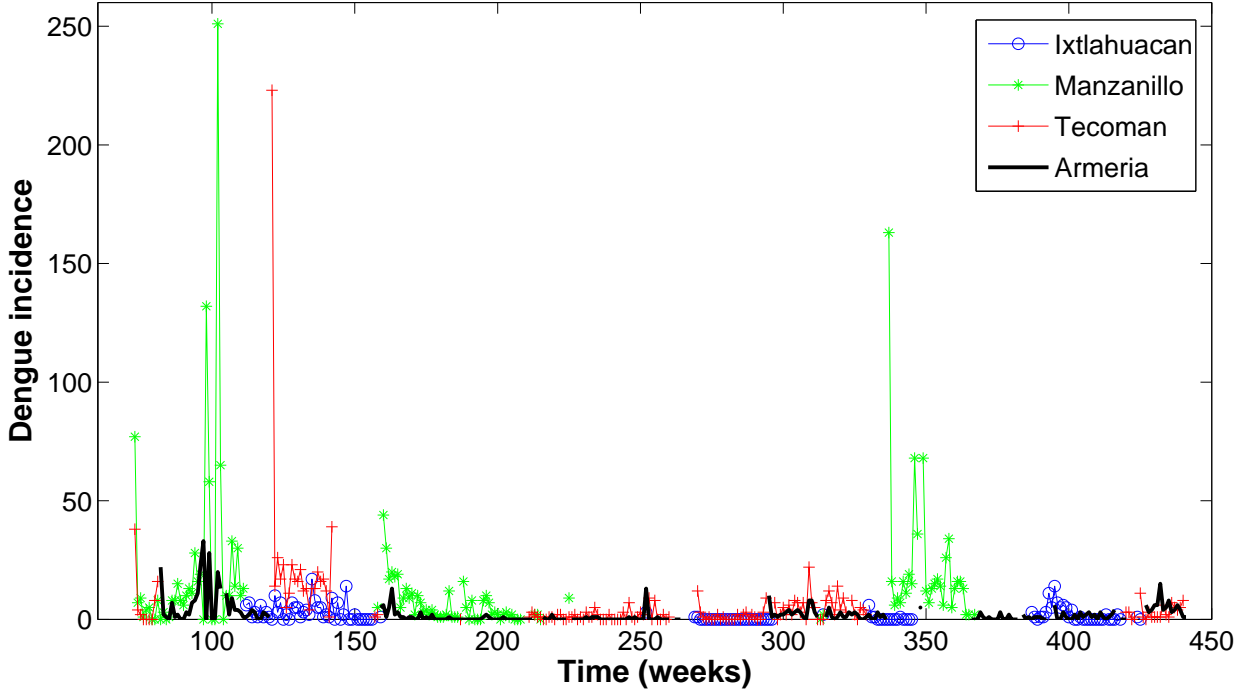}}
\caption{{\small Dengue incidence in funtion of the time at the four municipalities:  Armer\'ia, Ixtlahuac\'an, Manzanillo and Tecom\'an.}} \label{mati}
\end{center}
\end{figure}

In Fig. \ref{mati}, dengue incidence corresponding to the four municipalities addressed: Armer\'ia, Ixtlahuac\'an, Manzanillo and Tecom\'an between the years 2008 and 2015 is displayed.  \\

\begin{figure}[hbt] 
\begin{center}
\centerline{\includegraphics[width=1\textwidth]{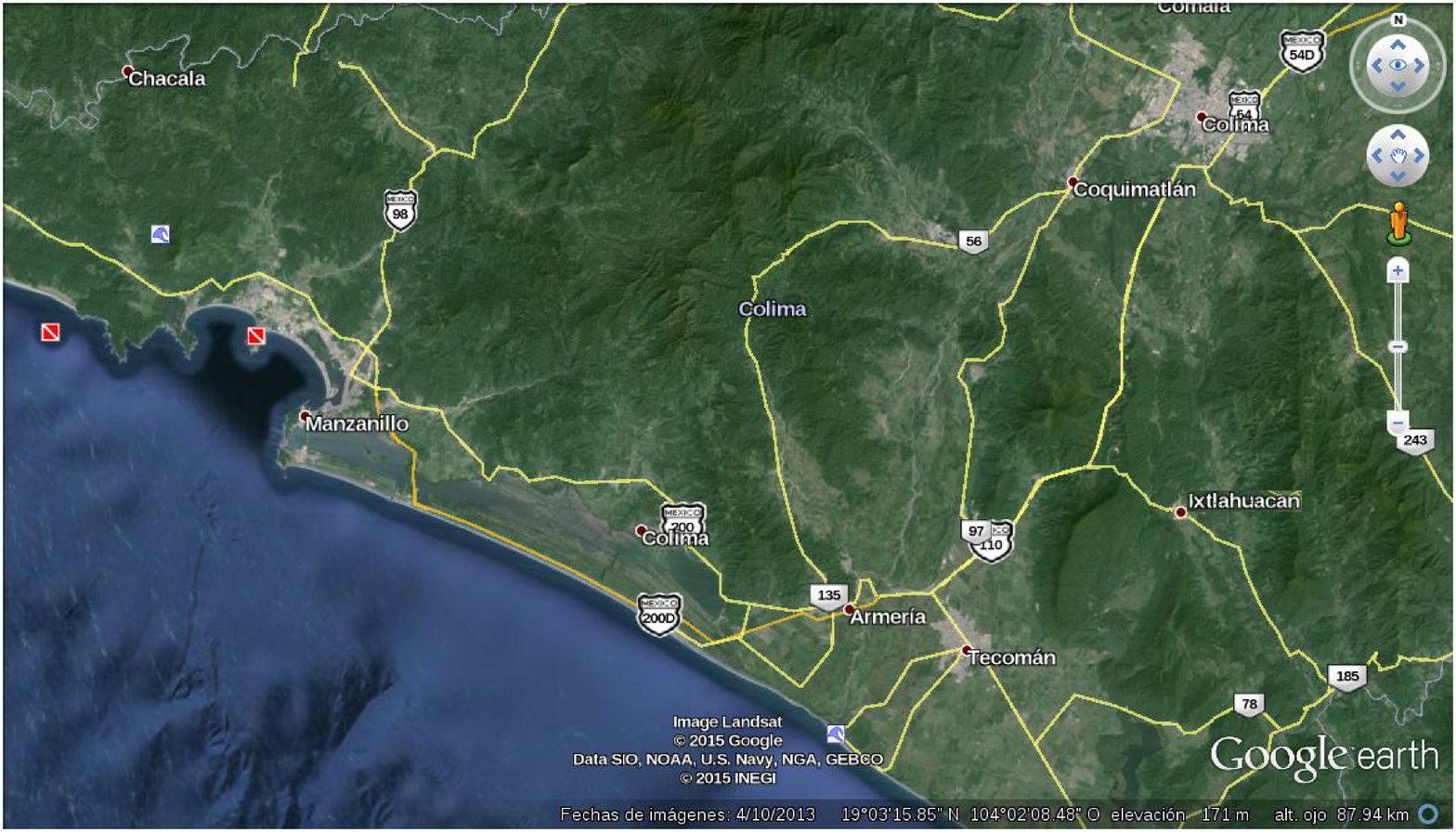}}
\caption{{\small Main highways in the Colima state. Picture taken from Google-Earth 2015.}} \label{caminos}
\end{center}
\end{figure}

The higher incidence began in Manzanillo, and it was propagated towards Armer\'ia, Tecom\'an and finally Ixtlahuac\'an. It is interesting to note that the main highways connecting Colima (the state capital) and Manzanillo, are traced in such a way that they necessarily pass through Ixtlahuac\'an, Tecom\'an, Armer\'ia to reach Manzanillo. We can see in Fig. \ref{caminos} the roads distribution in the region comprising the four municipalities of the state of Colima that we address. As it is apparent, there is not direct access from Colima to Manzanillo through Coquimatlan.  \\


\section{Conclusions} \label{se:concl}

Two demographically large municipalities (Manzanillo and Tecom\'an) and two small municipalities (Armer\'ia and Ixtlahuac\'an) were chosen. The former because they possess coastal regions and are economically and socially important. The latter because it emphasizes the fact that small municipalities with low population are also susceptible to high dengue incidence to the point of being considered red flag, in spite of being relatively isolated (Ixtlahuac\'an). This aspect certainly highlights the impact of the communication highways and human transit on dengue incidence for isolated areas.  \\

The graphs show a quasi-generalized pattern of dengue occurrence characterized by high values of the minimum temperature variable and precipitations. In fact, we observed that a high number of dengue cases reported in a particular municipality, which resulted in a red flag status was most commonly dictated by the temperature change. We can corroborate these as reported by \citet{Dc13} if we look at the graphs of figures \ref{arme}  to \ref{teco} in which all cases of dengue outbreaks occurred when the minimum temperature variable increased and during the presence of precipitation. \\

In Fig. \ref{mati} we can see that there is a wave of dengue incidence that is spatially spread throughout the four municipalities, starting in Manzanillo and ending towards Ixtlahuac\'an. Of course such spatial pattern does not finish there, but the study excluded municipalities that are located at the north of Ixtlahuac\'an. The data indicates that dengue propagates as an oriented wave, following the human transit (see Fig. \ref{caminos}) \citep{Mu92, Cu12}. \\

Although the number of dengue cases reported is low when compared to the size of the population in the municipalities, it is suggested to assign red flag status to a municipality with dengue incidence greater than ten cases to give track and prevent the spread, according with the DGE criterion \citep{Bo14}. \\

After DGE reported that Colima was the state with more dengue cases in 2009, the {\it descacharrizaci\'on} campaign was implemented. In the years in which such campaign was carried out, the dengue incidence decreased considerably, until no cases were reported for some periods of time. Unfortunately, this campaign was not continued and dengue incidences occurred again. This strongly indicates that the descacharrization campaign should have been performed periodically in each municipality to sensitize people in acquiring habits of order and cleanliness, especially with objects that are not thrown away and that represent an optimal place for mosquito's eggs incubation. \\

Weather data from official sites in Mexico are unfortunately not properly updated, which represents a drawback for studies targeting the effects of campaigns and public policies aiming at controlling dengue. However, by exploiting the relatively stable periodic behavior observed in climatological normals and the patterns found by applying the methodology of the present study for different municipalities, one could provide a prediction about the periodicity of dengue occurrence, and thus take appropriate measures to prevent it. \\

The methodology introduced in this report can be applied to other municipalities in the state of Colima and/or other states of Mexico that are considered regions with high prevalence of dengue, such as Yucatan, Jalisco, etc. In fact, determining not only temporal but also spatial predictive patterns of dengue occurrence in a specific region will be useful in designing public health policies that optimize economic resources while achieving better control strategies of dengue.    \\ 

Few cases of hemorrhagic dengue were reported in the municipalities that we addressed. Indeed, this number was considerable smaller than number of occurrences for classic dengue (dengue fever). By considering that hemorrhagic dengue has have a similar behavior than dengue fever [see \cite{Bo14}],  
we only focused on presenting results for dengue fever as being the most prevalent.

%

\bibliographystyle{natbib}


\end{document}